\documentclass[aps,prl,reprint,amsmath,amssymb,groupedaddress]{revtex4-2}
\usepackage[T1]{fontenc}
\usepackage{amsmath,amssymb,bm,mathtools}
\usepackage{graphicx}
\usepackage[hidelinks]{hyperref}

\newcommand{\MainEq}[1]{Eq.~(\ref{#1})}

\newcommand{\PaperTitle}{Quantum State Reconstruction from Sklar-Type Relational Data}
\newcommand{\PaperAbstract}{\begin{abstract}
Sklar's theorem in classical probability theory rigorously separates the marginal statistics of a joint probability distribution from its dependence structure data. We apply this perspective to the context-dependent classical probability tables induced by the Born rule for a quantum state. For each Born probability table of a given quantum state, we discard the marginal statistics of the individual parties, retaining only the relational data among them. Yet we prove that, when these classical relational data are collected over all contexts, the nontrivial quantum compatibility constraints inherent in quantum mechanics suffice to reconstruct any irreducible multiqubit quantum state itself. This means that the local information of the subsystems, discarded table by table, is fully recovered. We call this quantum-state reconstruction from Sklar-type relational data.
\end{abstract}     
}

\newcommand{\PaperAuthors}{%
  \author{Yi-Yu Lin}
  \email{yiyu@simis.cn}
  \affiliation{Fudan Center for Mathematics and Interdisciplinary Study, Fudan University, Shanghai 200433, China}
  \affiliation{Shanghai Institute for Mathematics and Interdisciplinary Sciences (SIMIS), Shanghai 200433, China}

  \author{Jie-Chen Jin}
  \affiliation{Department of Physics and Astronomy, University College London, Gower Street, London WC1E 6BT, U.K.}
}

\newcommand{\SuppRef}{Appendix~\ref{app:prelim}}
\newcommand{\AckHeading}{\section*{Acknowledgments}}
\begin{document}
\title{\PaperTitle}
\PaperAuthors
\PaperAbstract
\maketitle


\section{Introduction}\label{sec:intro}

For an entangled quantum state $\rho$, knowing the reduced states of all parties is still insufficient to determine the whole. This suggests that the relations among the parties themselves carry nontrivial information. Here we do not claim to invent an entirely new mathematical concept of quantum relations.
Rather, we merely aims to call attention to the fact that the Born rule already provides a simple entry point: once a set of local measurement bases, i.e., a \emph{context}, is chosen, then a state $\rho$ corresponds to a joint probability table; and as the context changes, the table changes accordingly. 
Within a fixed context, however, this gives merely an ordinary classical probability table, with no mathematical difference from the objects studied by statisticians. 
That being so, why not borrow their mathematical tools to study the relations therein? It turns out that the famous Sklar theory in classical probability theory \cite{Sklar1959,Nelsen2006} has provided us with a valuable perspective:
 in a joint probability distribution, the marginal statistics of the individual variables can be strictly separated from the dependence structure among them. 
In other words, in a certain sense, the purely relational structure can be studied as an independent mathematical object, provided that we strictly discard the marginal data of the individual parties. 

In this Letter, we applying this procedure to the Born tables generated by a multi-qubit quantum state in different contexts, thereby obtaining an entire family of relational data, which we call Sklar-type relational data. 
Remarkably, we find that this entire family of relational data can in fact conversely determine the original quantum state, as long as it admits no further tensor-product factorization.  
The secret lies in the fact that the Born tables in different contexts are subject to the nontrivial quantum compatibility constraint that they must all arise from the same parent density operator. 
In this sense, the simple intuition about entanglement and the whole even acquires a deeper meaning: under quantum compatibility constraints, all the relational manifestations of a quantum state almost already determine the quantum state itself.

It is worth mentioning that the idea of appreciating quantum mechanics from a relational perspective has deep historical roots; see, e.g., \cite{schrodinger1935probability,teller1986holism,rovelli1996relational,mermin1998quantum,esfeld2004relations}.
In addition, some previous works have explored connections between Sklar's theorem or copula theory and quantum mechanics in various aspects, such as \cite{aladilee2009copula, lovas2019quantumcopulas, zhu2022generative, zhu2023copularisk}.
However, the research objective considered here is, in a sense, closer to quantum state tomography \cite{dariano2003tomography,james2001measurement}.  The difference is that traditional tomography usually reconstructs a quantum state using informationally complete measurement statistics, whereas we investigate whether sufficiently rich relational data can still possess reconstructive power even after stripping away the single-party marginal statistics from each joint probability table and retaining only the dependence structure.

The paper is organized as follows. We first review the classical bivariate discrete Sklar decomposition and apply it to the reconstruction of two-qubit quantum states. After extending Sklar decomposition to the multivariate setting, we establish a three-qubit reconstruction theorem and reveal in detail the specific conditioning techniques and cross-context compatibility mechanism involved.  
We then use a recursive method to extend this mechanism to general $N$-qubit systems. Some supplementary technical details are given in \SuppRef.

\section{Bivariate Sklar decomposition review}
\label{sec:bivariate-sklar}

This letter focuses on multi-qubit systems. To extract relational data characterizing the dependence structure of their Born probability tables independently of single-party marginals, we adopt a mathematical construction from classical contingency-table statistics \cite{bishopFienbergHolland1975,yule1900,yule1912,mosteller1968}: joint probability tables differing only by local positive reweighting are regarded as having the same dependence, and their equivalence class is taken as the relational object. 
The resulting dependence equivalence class is called the \emph{dependence nucleus}, denoted by $[p]$ \cite{geenens2020,geenens2025}. This construction embodies precisely the spirit of the classical Sklar theorem \cite{Sklar1959}: the marginal statistics and dependence structure of a joint distribution can be strictly separated, and after being specified separately, they can uniquely re-determine the entire joint distribution. Their mathematical correspondence is summarized in Table~\ref{tab:sklar}. A more detailed account of this correspondence and of the standard Sklar theorem itself is given in \SuppRef.

Consider the simplest bivariate binary case. Let $p=(p_{st})$ be a strictly positive joint probability table, where $s,t=\pm1$. Here, local reweighting amounts simply to multiplying each row and column by positive factors and then renormalizing:
\begin{equation}
p'_{st}=\frac{a_s b_t p_{st}}{Z},
\qquad a_s,b_t>0.
\label{eq:loc}
\end{equation}
Moreover, in this simple case, the seemingly abstract relational equivalence class $[p]$ can in fact be completely characterized by a single number, which is called the \emph{odds ratio}  in traditional contingency-table theory \cite{yule1900}:
\begin{equation}
\omega(p)=\frac{p_{++}p_{--}}{p_{+-}p_{-+}}.
\label{eq:odds}
\end{equation}
Under the local reweighting \eqref{eq:loc}, the weight factors $a_s,b_t$ cancel out exactly in $\omega$. More importantly, the converse also holds: for any two strictly positive bivariate binary tables $p,q$, 
\begin{equation}
\omega(p)=\omega(q)
\quad\Longleftrightarrow\quad
[p]=[q].
\label{eq:char}
\end{equation}
Therefore, $\omega$ is not only an invariant, but also completely characterizes the relational structure defined by the aforementioned reweighting. In particular, $\omega=1$ if and only if the two parties are statistically independent. 

\begin{table*}[t]
\caption{\label{tab:sklar}
Correspondence between discrete relational structures and classical continuous Sklar theory.
}
\begin{ruledtabular}
\begin{tabular}{p{0.19\textwidth}p{0.31\textwidth}p{0.31\textwidth}}
Mathematical structure
& Bivariate binary case (this work)
& Standard continuous Sklar theory \\
\hline
Sklar-type decomposition
& $p\longleftrightarrow(p_X,p_Y,[p])$
& $F\longleftrightarrow(F_X,F_Y,C)$ \\

Abstract dependence object
& Dependence nucleus $[p]$
& Equivalence class of joint distributions sharing the same copula \\

Complete characterization of dependence
& Odds ratio $\omega(p)$
& Copula function $C$
\end{tabular}
\end{ruledtabular}
\end{table*}

\section{Basic Picture and Two-Qubit State Reconstruction}
\label{sec:two-qubit}

Now consider a two-qubit quantum state $\rho_{AB}$. 
Once a set of local measurement bases is chosen on each party, i.e., specifying a context, the Born rule yields the probabilities of the four possible joint measurement outcomes. More explicitly, we first write $\rho_{AB}$ in the standard Fano--Bloch form \cite{Fano1983,SchlienzMahler1995}:
\begin{equation}
\begin{aligned}
\rho_{AB}=\frac{1}{4}\bigg[
&I\otimes I
+(\mathbf{a}\cdot\boldsymbol{\sigma})\otimes I\\
&+I\otimes(\mathbf{b}\cdot\boldsymbol{\sigma})
+\sum_{i,j}T_{ij}\sigma_i\otimes\sigma_j
\bigg].
\end{aligned}
\label{eq:fano}
\end{equation}
Here, $\mathbf{a}$ and $\mathbf{b}$ record the local spin expectation values of the two qubits, respectively, while the matrix $T$ records the expectation values of the products of their Pauli measurement outcomes. 
From Eq.~\eqref{eq:fano}, it is clear that once the parameters $(\mathbf{a},\mathbf{b},T)$ are determined,  the quantum state $\rho_{AB}$ is completely determined. 
Let $\mathbf{n}$ and $\mathbf{m}$ denote the unit directions on the Bloch sphere along which the two qubits are measured, with outcomes $s,t=\pm1$. Then applying the Born rule to Eq.~\eqref{eq:fano}, we obtain the following probability table for the outcomes $(s,t)$:
\begin{equation}
p_{st}^{\rho}(\mathbf{n},\mathbf{m})
=\frac{1}{4}\left[
1+s\,\mathbf{a}\cdot\mathbf{n}
+t\,\mathbf{b}\cdot\mathbf{m}
+st\,\mathbf{n}^{T}T\mathbf{m}
\right].
\label{eq:pst}
\end{equation}
Thus, for each fixed context $(\mathbf{n},\mathbf{m})$, Eq.~\eqref{eq:pst} is nothing but an ordinary $2\times2$ joint probability table---precisely the statistical object discussed in the previous section. As the measurement context varies, the same quantum state $\rho_{AB}$ generates an entire family of such classical joint probability tables, which we shall henceforth call \emph{Born tables}.
\footnote{ In fact, the discrete Sklar-type decomposition adopted in this paper are all aimed at joint probability tables whose entries are all strictly positive. Therefore, this paper only uses the relational data provided by Born tables with strictly positive probabilities (i.e., \emph{regular contexts}).
We deliberately discard all Born tables containing zero probabilities, and do not exploit the additional information about which measurement directions produce such zeros. This is because this kind of information can sometimes reveal the local state of the subsystems themselves, which goes against our original intent of studying only the relations among the parties.  
Moreover, unlike our earlier work \cite{Lin:2026meu}, which proves things using a bounded dependence quantity $Q$ defined over all contexts, we further show that relational data from regular contexts alone are already sufficient to accomplish reconstruction, including the cases of pure states and rank-deficient states.  Mathematically, this treatment benefits from the analyticity of Born probabilities with respect to measurement contexts: identities established from relational data on regular contexts can be converted into denominator-free algebraic identities regarding Born probabilities, and then analytically continued to the entire measurement space. Strict arguments in this regard can be found in \SuppRef.}
Now, as outlined in the Introduction, we investigate the relational structure of each Born table. 
The preceding Sklar-type decomposition tells us that, upon intentionally discarding the marginal statistics of each Born table, the strictly isolated odds ratio precisely characterizes the dependence relation between the two parties. Applying this treatment one by one to the Born tables of the same quantum state in different contexts, we obtain an entire family of purely relational data, $\omega_{\rho_{AB}}(\mathbf{n},\mathbf{m})$. 
And our unexpected discovery is that, in quantum mechanics, this entire family of purely relational data is surprisingly sufficient to conversely determine the state of the whole $\rho_{AB}$! 
This also means that the local marginal information, originally discarded on purpose just study relations in their own right, is almost entirely recovered when the full family of Born contexts is taken into account. More precisely, we have the following theorem: 

\medskip
\noindent\textbf{Theorem 1.---}
For any non-product two-qubit quantum state $\rho_{AB}$, the relational data of its classical Born tables over all local measurement contexts $(\mathbf{n},\mathbf{m})$, i.e., the odds-ratio family
$\{\omega_{\rho_{AB}}(\mathbf{n},\mathbf{m})\}$,
are sufficient to reconstruct the original state, up to a discrete ambiguity.
\footnote{More precisely, the two candidate states possessing the same all-context odds-ratio family are $\rho$ and
$ \protect\widetilde{\rho}{\rho}
=(\sigma_y\otimes\sigma_y)\rho^*
(\sigma_y\otimes\sigma_y)$.
This discrete ambiguity can be understood intuitively as a freedom in the convention for the outcome labels: simultaneously exchanging the labels $+$ and $-$ on both sides in every measurement context gives
$p_{st}^{ \protect\widetilde{\rho}{\rho}}(\mathbf{n},\mathbf{m})
=p_{-s,-t}^{\rho}(\mathbf{n},\mathbf{m})$,
while leaving the odds ratio of every Born table unchanged. Hence the relational data alone cannot distinguish these two choices.}
\medskip

This reconstruction result has been rigorously proven in our earlier work \cite{Lin:2026meu}, and we will not repeat the details of its proof here. Below, we discuss only the basic idea behind it. 
The key point is that all Born contexts provide us with an family of sufficiently rich relational data, enabling us to always find, for different candidate quantum states, some specific discriminative contexts that serve as clues ultimately sufficient to reconstruct the state $\rho_{AB}$. 
More explicitly, it can be proven that the all-context relational data family $\{\omega_{\rho_{AB}}(\mathbf{n},\mathbf{m})\}$ forms a \emph{separating family}. Namely, for any candidate state $\sigma$, characterized by a parameter triple $(\mathbf{a}',\mathbf{b}',T')$, that attempts to masquerade as $\rho_{AB}$, one can extract from this family special contexts sufficient to distinguish the two and expose it:.  if $\sigma$ possesses exactly the same relational data as $\rho_{AB}$, then their parameters $(\mathbf{a},\mathbf{b},T)$ and $(\mathbf{a}',\mathbf{b}',T')$ must either match perfectly, or differ only by the discrete ambiguity stated in the theorem (i.e., $(\mathbf{a}',\mathbf{b}',T')=(-\mathbf{a},-\mathbf{b},T)$).

\section{Multivariate Sklar Decomposition Review}
\label{sec:multivariate-sklar}

We continue our review of the Sklar-type decomposition, now considering the characterization of relational data for strictly positive joint distributions $p(x)$ of $N$ binary random variables $x_i=\pm1$. The local positive reweighting in Eq.~\eqref{eq:loc} is now naturally generalized to
\begin{equation}
q(x)=\frac{1}{Z}p(x)\prod_{i=1}^{N}f_i(x_i),
\qquad f_i>0.
\label{eq:multi-loc}
\end{equation}
The resulting dependence equivalence class is still denoted by $[p]$. In the bivariate case, a single odds ratio \eqref{eq:odds} suffices to characterize $[p]$ completely; for the multipartite case, however, an entire set of relational data is required. To this end, we adopt the standard log-linear expansion from classical contingency-table theory \cite{good1963,goodman1964,buchman2012}: 
\begin{equation}
\begin{gathered}
\log p(x)=\sum_{S\subseteq[N]}J_S\prod_{i\in S}x_i,\\
[N]=\{1,\ldots,N\}.
\end{gathered}
\label{eq:log-linear}
\end{equation}
As rigorously established in Ref.~\cite{Lin:2026trf}, the complete characterization provided by the odds ratio in the bivariate case, Eq.~\eqref{eq:char}, now generalizes to
\begin{equation}
[p]=[q]\quad\Longleftrightarrow\quad
J_S(p)=J_S(q),\quad \forall\,|S|\geq2.
\label{eq:multi-dep}
\end{equation}
That is, the coefficients $J_S$ of second and higher orders ($|S|\geq2$) precisely and completely characterize the multivariate dependence nucleus $[p]$. 
Moreover, these relational data can be specified independently of arbitrary strictly positive single-party marginal distributions, with each such combination uniquely determining the entire joint distribution. In particular, when $N=2$, the sole relational coordinate $J_{\{1,2\}}$ is simply
\begin{equation}
J_{\{1,2\}}=\frac{1}{4}\log\omega,
\end{equation}
which precisely returns to the previous odds-ratio description \eqref{eq:odds}. 
In fact, these coefficients $J_S$ in Eq.~\eqref{eq:multi-dep} admit a rather direct statistical-mechanical interpretation \cite{Lin:2026trf}: for a classical many-spin Gibbs distribution, they correspond exactly to the dimensionless couplings $J_S=\beta K_S$ of two-body and higher-order interactions, whereas local reweighting \eqref{eq:multi-loc} amounts to changing the external fields acting on individual spins without altering the interactions among them.

\section{Three-Qubit State Reconstruction}
\label{sec:three-qubit}

Now consider a three-qubit quantum state $\rho_{ABC}$. In each local measurement context, the Born rule first yields an ordinary three-variable joint distribution $p_{str}$, where $s,t,r\in\{\pm1\}$ denote the measurement outcomes of $A,B,C$, respectively. For this $2\times2\times2$ joint probability table, a natural idea is to perform the following standard statistical treatment \cite{Agresti2013}: fix the outcome of one variable, say $C=r$, and consider the resulting conditional $2\times2$ table:
\begin{equation}
p_{st\mid r}=\frac{p_{str}}{p_r},
\qquad
p_r=\sum_{s,t}p_{str},
\label{eq:fixr}
\end{equation}
and then define its conditional odds ratio,
\begin{equation}
\begin{split}
\omega_{AB\mid C=r}
&=\frac{p_{++\mid r}p_{--\mid r}}
{p_{+-\mid r}p_{-+\mid r}}\\
&=\frac{p_{++r}p_{--r}}
{p_{+-r}p_{-+r}}.
\end{split}
\label{eq:wabc}
\end{equation}
Notice that the probability $p_r$ of the conditioning event itself does not enter the dependence structure of this conditional table. 
Yet this ordinary statistical treatment provides us a rather ingenious opportunity. Once the measurement direction and outcome of the third party are fixed, as the measurement contexts of the other two parties vary arbitrarily, the resulting entire family of conditional Born probabilities can, remarkably, be represented by a single two-qubit density operator! 
In fact, if we define the following normalized state operator on $AB$:
\begin{equation}
\begin{split}
\widehat{\tau}_{AB\mid C,r}^{\rho}(\mathbf n_C)
=\frac{1}{p_r^{\rho}(\mathbf n_C)}
\operatorname{Tr}_C\bigl[
&(I_{AB}\otimes\Pi_r^C)\rho_{ABC}\\
&\times(I_{AB}\otimes\Pi_r^C)\bigr],
\end{split}
\label{eq:tao}
\end{equation}
where
\begin{equation}
\Pi_r^C(\mathbf n_C)
=\frac12\bigl(I_C+r\,\mathbf n_C\cdot\boldsymbol\sigma^C\bigr)
\label{eq:proj}
\end{equation}
represents the projection operator corresponding to outcome $r=\pm1$ when $C$ is measured along the Bloch direction $\mathbf n_C=(n_x,n_y,n_z)$. 
Then, for all local measurements on $AB$, it exactly reproduces the conditional Born probabilities $p_{st\mid r}$ defined previously: 
\begin{equation}
\operatorname{Tr}\!\left[
(\Pi_s^A\otimes\Pi_t^B)
\widehat{\tau}_{AB\mid C,r}^{\rho}(\mathbf n_C)
\right]
=p_{st\mid r}.
\label{eq:born}
\end{equation}
In other words, $\widehat{\tau}_{AB\mid C,r}^{\rho}(\mathbf n_C)$ provides a unified mathematical quantum-state representation of this entire family of conditional Born probabilities, while $\omega_{AB\mid C=r}$ exactly constitutes its complete two-qubit all-context dependence data. 
In this way, as long as this two-qubit operator is not a product state, we have the opportunity to directly apply the previous two-qubit reconstruction theorem,  and thereby attempt to recover the original three-qubit state. Following this line of thought, we can prove the following theorem:

\medskip
\noindent\textbf{Theorem 2.---}
For any three-qubit quantum state $\rho_{ABC}$ with no isolated qubit, the dependence structures of its Born distributions over all local measurement contexts suffice to uniquely reconstruct the quantum state.
\medskip

Here ``no isolated qubit'' means that none of the partitions $A\mid BC$, $B\mid AC$, and $C\mid AB$ admits a tensor-product factorization.

Returning to the reconstruction argument, we must first check whether $\widehat{\tau}_{AB\mid C,r}^{\rho}(\mathbf n_C)$ might happen to be a product state. The answer is that special measurement directions can indeed yield produce conditional states. 
However, since these conditional states are not generated independently, but are continuously and linearly generated from the same three-qubit state as the measurement direction of $C$ varies, it can be rigorously proved that exceptive product conditional states pose no obstacle.
\footnote{ Simply put, the genuinely problematic case would be productness persisting throughout an open region of conditioning directions. Such persistent factorization, however, forces a common tensor factor, which in turn implies that the parent state already has a product cut. A rigorous proof is given in \protect\SuppRef.} 
Now let $\sigma_{ABC}$ be another candidate state possessing the same all-context dependence data. Theorem~1 then tells us that
\begin{equation}
\begin{gathered}
\widehat{\tau}_{AB\mid C,r}^{\sigma}(\mathbf n_C)
=\widehat{\tau}_{AB\mid C,r}^{\rho}(\mathbf n_C),\\
\text{or}\\
\widehat{\tau}_{AB\mid C,r}^{\sigma}(\mathbf n_C)
=\mathcal F_A\mathcal F_B\!\left[
\widehat{\tau}_{AB\mid C,r}^{\rho}(\mathbf n_C)
\right].
\end{gathered}
\label{eq:or}
\end{equation}
Thus far, the full all-context dependence data have determined the normalized $AB$ conditional state under each conditioning context up to the simultaneous double-spin-flip ambiguity.
\footnote{Strictly speaking, for each measurement direction ${\mathbf n_C }$ of $C$, the two-qubit theorem only tells us that the corresponding conditional $AB$ state belongs to either the identity branch or the double-spin-flip branch. Due to the continuous dependence of the conditional state on ${\mathbf n_C }$, it can be proven that one of the branches must consistently hold within some nonempty open region of directions for measuring $C$. A rigorous proof is given in \protect\SuppRef.}
Furthermore, using the Pauli operators $\{I,\sigma_x,\sigma_y,\sigma_z\}$ as a basis for the operator space on qubit $C$, the state $\rho_{ABC}$ has the following unique expansion \cite{NielsenChuang2010}: 
\begin{equation}
\rho_{ABC}=\frac12\left[
M_0^{AB}\otimes I_C
+\sum_{\mu=x,y,z}M_\mu^{AB}\otimes\sigma_\mu^C
\right].
\label{eq:exp}
\end{equation}
In other words, once the four coefficient operators $M_0^{AB},M_x^{AB},M_y^{AB},M_z^{AB}$ are known, $\rho_{ABC}$ is completely determined. The key is that these Pauli expansion coefficients are closely related to the quantities obtained by statistical conditioning:
\begin{equation}
\begin{split}
p_r^{\rho}(\mathbf n_C)
\widehat{\tau}_{AB\mid C,r}^{\rho}(\mathbf n_C)
=\frac12\biggl[
M_0^{AB}
+r\sum_{\mu=x,y,z}n_\mu M_\mu^{AB}
\biggr].
\end{split}
\label{eq:con}
\end{equation}
The point of Eq.~\eqref{eq:con} is that, although the conditioning probabilities $p_r^{\rho}(\mathbf n_C)$ have been eliminated from the conditional odds ratio \eqref{eq:wabc}, they must still, together with the conditional density-operator representations at different measurement directions, jointly determine the same set of Pauli expansion coefficients $M_\mu^{AB}$ of the same parent state, according to  
\begin{equation}
\begin{aligned}
M_0^{AB}={}&p_+^{\rho}(\mathbf e_x)
\widehat{\tau}_{AB\mid C,+}^{\rho}(\mathbf e_x)\\
&+p_-^{\rho}(\mathbf e_x)
\widehat{\tau}_{AB\mid C,-}^{\rho}(\mathbf e_x),
\end{aligned}
\label{eq:read1}
\end{equation}
\begin{equation}
\begin{aligned}
M_\mu^{AB}={}&p_+^{\rho}(\mathbf e_\mu)
\widehat{\tau}_{AB\mid C,+}^{\rho}(\mathbf e_\mu)\\
&-p_-^{\rho}(\mathbf e_\mu)
\widehat{\tau}_{AB\mid C,-}^{\rho}(\mathbf e_\mu),\\
&\hspace{2cm}\mu=x,y,z.
\end{aligned}
\label{eq:read2}
\end{equation} 
Below we show that the cross-context constraint \eqref{eq:con} indeed allows us to determine the probabilities $p_r^{\rho}(\mathbf n_C)$. 
Let the Pauli expansion coefficients of $\sigma_{ABC}$ parallel to those in Eq.~\eqref{eq:exp} be $N_\mu^{AB}$. 
Then, using Eq.~\eqref{eq:con} together with Eq.~\eqref{eq:or}, we further obtain, on the identity branch,
\begin{equation}
\begin{split}
p_r^{\sigma}(\mathbf n_C)\biggl[
M_0^{AB}+r\sum_\mu n_\mu M_\mu^{AB}\biggr]
= {} &\\
p_r^{\rho}(\mathbf n_C)\biggl[
N_0^{AB}+r\sum_\mu n_\mu N_\mu^{AB}\biggr].
\end{split}
\label{eq:squa}
\end{equation}
Equation~\eqref{eq:squa} is in fact an operator-valued polynomial identity of degree at most two in $\mathbf n_C$.  By comparing coefficients of different orders, one can prove\footnote{The rigorous coefficient comparison and the treatment of degenerate cases under the Bloch-sphere constraint are given in \protect\SuppRef.} that, unless the normalized conditional state is completely independent of $\mathbf n_C$, necessarily
\begin{equation}
p_r^{\sigma}(\mathbf n_C)=p_r^{\rho}(\mathbf n_C),
\label{eq:wise}
\end{equation}
and consequently
\begin{equation}
N_\mu^{AB}=M_\mu^{AB}.
\label{eq:nm}
\end{equation} 
The former degenerate case would precisely force $C$ to be in a product state with $AB$, which has already been excluded by the theorem's assumption. 
For the simultaneous double-spin-flip branch of Eq.~\eqref{eq:or}, the argument is entirely parallel, since $\mathcal F_A\mathcal F_B$ is real-linear and trace-preserving. In summary, from the perspective of choosing qubit $C$ as the subsystem on which we condition, we have so far proved  
\begin{equation}
\sigma_{ABC}=\rho_{ABC}
\quad\text{or}\quad
\sigma_{ABC}=\mathcal F_A\mathcal F_B(\rho_{ABC}).
\label{eq:branchC}
\end{equation} 
The same argument can of course be carried out by conditioning on either of the other two qubits respectively. Therefore, the same candidate parent state $\sigma_{ABC}$ must in fact be simultaneously compatible with all three conditioning decompositions. 
It turns out that this further eliminates even the residual freedom of relabeling local spin outcomes that could not be distinguished in the two-qubit case. This follows from a simple reductio ad absurdum.  Hence
\begin{equation}
\sigma_{ABC}=\rho_{ABC}.
\end{equation}
The theorem is proved.

\section{The Sklar Reconstruction Mechanism for Quantum States}
\label{sec:sklar-mechanism}

Recall from the preceding multivariate Sklar decomposition that the relational data of a multipartite probability table are completely characterized by its higher-order coefficients $J_S$. Yet these coefficients seemed not to appear in the three-qubit reconstruction story just now. In fact, they have always been the main protagonists participating in the reconstruction, just in a slightly different guise. Actually, the conditional odds ratio \eqref{eq:wabc} satisfies
\begin{equation}
\log\omega_{AB\mid C=r}
=4\left(J_{AB}+rJ_{ABC}\right).
\label{eq:conditional-odds}
\end{equation} 
This means that the conditional relational data we have just used are precisely combinations of the original two- and three-body relational data. Of course, performing conditioning on a classical probability table is, in itself, merely a standard statistical treatment. The key is that something mathematically remarkable happens next: Eqs.~\eqref{eq:tao} and \eqref{eq:born} show that, as the measurement contexts of the remaining two parties vary arbitrarily, the resulting entire family of conditional probability tables admits a particularly simple Hilbert-space operator representation---namely, it can be equivalently described by applying the Born rule to a single two-qubit density operator. The ingenuity lies in the fact that this mathematical representation makes Theorem~1, the lower-body quantum-state reconstruction theorem based solely on relational data, directly applicable. 

This success suggests a lesson worth generalizing. Facing a more general joint probability table of $N$ binary variables, we may likewise first fix the outcome $x_k=r$ of the $k$th variable to obtain a conditional probability table of $N-1$ variables. 
Unlike the $N=3$ case, these new conditional tables are now generally still multivariate, so their relational data must also be characterized by a set of coefficients $J_T$. It turns out that all their relational coordinates satisfy
\begin{equation}
\begin{gathered}
J_T^{(k,r)}
=J_T+rJ_{T\cup\{k\}},\\
|T|\geq2,\qquad k\notin T.
\end{gathered}
\label{eq:conditional-J}
\end{equation}
Similar to Eq.~\eqref{eq:conditional-odds}, this is simply a recombination of the original global relational data.\footnote{A rigorous proof of this identity is given in \protect\SuppRef.} The key is that Eqs.~\eqref{eq:tao} and \eqref{eq:born} can clearly be straightforwardly generalized to the general case where the remaining system comprises more than two qubits. Consequently, the family of conditional Born probabilities still admits a unified representation by a lower-body density operator, which gives us confidence that the reconstruction mechanism just revealed can be implemented in a recursive manner.

We defer the rigorous mathematical confirmation that this mechanism indeed yields a recursive reconstruction of general multiqubit states to the next section. 
Before that, let us examine the reconstruction mechanism itself a little more closely. After all, it answers precisely the question of greatest interest in this letter: how can the entire set of the classical relational manifestations of a quantum state possibly suffice to determine the quantum state itself?
 In other words, how can the marginal information discarded in the classical Sklar decomposition nevertheless be recovered from these relational data? 
First, let us revisit Eq.~\eqref{eq:fixr}. Obviously, once the lower-body reconstruction theorem has determined $\widehat{\tau}^{\rho}$, we need only determine the conditioning weights $p_r$---which are themselves the marginal probabilities of the party being conditioned on---to recover the original joint Born probability table, thereby automatically recovering the marginal statistics of the other parties as well. 
The remarkable thing lies in the fact that these weight coefficients $p_r$ cannot be chosen arbitrarily. 
Admittedly, in the ordinary statistical setting of a single fixed context, as marginal data, they can indeed be freely specified within the classical Sklar decomposition. 
However, as indicated by Eq.~\eqref{eq:con}, these data must actually satisfy a very stringent requirement: 
the conditional statistics from different measurement contexts can arise from a single parent quantum state only when combined with particular probability weights. 
And what Eq.~\eqref{eq:squa} further does is precisely to further demonstrate that this requirement is strong enough to fix those weights.

\section{N-Qubit State Reconstruction}
\label{sec:n-qubit}

We can now state the general reconstruction theorem.

\medskip
\noindent\textbf{Theorem 3.---}
For any $N\geq3$, an $N$-qubit quantum state $\rho$ that admits no tensor-product factorization across any nontrivial partition can be uniquely reconstructed from the relational data of its Born distributions over all local measurement contexts.\footnote{For quantum states admitting nontrivial tensor-product factorizations, it suffices to consider their irreducible components separately. }
\medskip

The proof simply requires applying the previous mechanism recursively. The $N=3$ case has already been established by Theorem~2. 
For $N\geq4$, let $\rho$ and $\sigma$ possess the same all-context dependence data. Choose the $k$th qubit for conditioning, i.e., fix its measurement direction $\mathbf n_k$ and outcome $r$. 
By Eq.~\eqref{eq:conditional-J}, the conditional Born probability families of the remaining $N-1$ parties have the same relational data for both states.
For generic conditioning directions, their unified density-operator representations remain irreducible under tensor-product factorization (a rigorous proof is given in \SuppRef). The induction hypothesis therefore yields 
\begin{equation}
\widehat{\tau}_{\bar{k}\mid k,r}^{\sigma}(\mathbf n_k)
=
\widehat{\tau}_{\bar{k}\mid k,r}^{\rho}(\mathbf n_k).
\label{eq:N-tau}
\end{equation}
Here $\bar{k}$ denotes the remaining $N-1$ qubits. In particular, since $N-1\geq3$, the double-spin-flip ambiguity of two-qubit reconstruction no longer arises.
It remains only to recover the conditioning probabilities. Note that Eq.~\eqref{eq:con} from the three-qubit version proof still holds here:
\begin{equation}
\begin{aligned}
&p_r^{\rho}(\mathbf n_k)
\widehat{\tau}_{\bar{k}\mid k,r}^{\rho}(\mathbf n_k)\\
&\qquad=\frac12\left[
M_0^{\bar{k}}
+r\sum_{\mu=x,y,z}n_{k,\mu}M_\mu^{\bar{k}}
\right].
\end{aligned}
\label{eq:N-con}
\end{equation}
Although the four operators $M_\mu^{\bar{k}}$ now act on a larger Hilbert space, the linear structure of Eq.~\eqref{eq:N-con} with respect to $\mathbf n_k$ remains entirely unchanged. 
Therefore, the cross-context compatibility argument, exactly parallel to that for Eq.~\eqref{eq:squa}, still establishes that the conditioning probabilities of the two candidate states must coincide. The only possible degenerate case would imply that the $k$th qubit was isolated from the rest from the outset, which is excluded by assumption (see \SuppRef for details). Consequently, all the Pauli expansion coefficients in Eq.~\eqref{eq:N-con} must also coincide, giving $\sigma=\rho$ and completing the induction.

\section{Conclusion and Outlook}
\label{sec:conclusion}

In this work, we have simply used the Born rule to characterize a specific quantum state by a family of context-dependent classical probability tables. 
We then employed the classical Sklar theory to turn the idea of ``retaining only relational data'' for each classical probability table into a rigorous mathematical procedure. 
This provides a route to quantum-state reconstruction distinct from directly collecting all local statistics: starting solely from the relational data of a family of classical probability tables, their common underlying density operator can be uniquely identified through the quantum compatibility constraints among them. Of course, this method requires the quantum state itself to form a relational block admitting no further tensor-product factorization. Yet, in the spirit of Sklar theory, when two systems have no relations at all, there is no reason to force them into a single relational whole in the first place. 

We conclude with several open directions. The first concerns potential experimental applications. if reconstruction schemes can be further developed that require only a limited measurement context, and to be robust against statistical errors and computationally feasible, then quantum state reconstruction using purely relational data may become a complementary technique to conventional quantum-state tomography. In particular, the inherent insensitivity of the Sklar-type relational data used here to local positive reweighting, which preserves the dependence structure, may offer a distinctive advantage for reconstructing quantum states without precise knowledge of the relative detection efficiencies of individual local measurement outcomes. 
Another natural direction is to extend Sklar-type relational reconstruction to continuous-variable quantum systems. In particular, for multimode quantum states probed by quadrature measurements, although complete quadrature statistics already suffice for quantum-state tomography \cite{vogel1989quadrature,lvovsky2009tomography}, whether the original state can still be reconstructed after further discarding the local marginals in each quadrature context and retaining only the corresponding Sklar copula remains an open question worth investigating.

\AckHeading
\begin{acknowledgments}
This work was supported by the National Natural Science Foundation of China (NSFC) under Grant No. 1250050230.
We thank ChatGPT (OpenAI, GPT-6) for valuable assistance with technical details of the mathematical proofs, particularly those in the Supplemental Material, and with English-language polishing. The AI-assisted work was directed and reviewed by the authors, who take full responsibility for the final manuscript.
\end{acknowledgments}

\clearpage
\appendix
\setcounter{secnumdepth}{1}

\section{Classical Sklar Structures and Conditioning}
\label{app:prelim}

Since we do not assume that physicists are familiar with Sklar's theorem, we explain it here in somewhat greater detail for self-containment.

\subsection{Review of Continuous Sklar Theory and Its Structural Correspondence with the Discrete Dependence Nucleus}

The standard formulation of Sklar's theorem \cite{Sklar1959} is as follows. Let $F$ be an $N$-dimensional joint cumulative distribution function (CDF) with marginal distributions $F_1,\ldots,F_N$. Then there exists an $N$-dimensional copula $C$ such that
\begin{equation}
F(x_1,\ldots,x_N)
=C\bigl(F_1(x_1),\ldots,F_N(x_N)\bigr).
\label{eq:sm-sklar}
\end{equation}
If the marginals are continuous, $C$ is unique.\footnote{More precisely, $C$ is unique on $\mathrm{Ran}(F_1)\times\cdots\times\mathrm{Ran}(F_N)$.} Conversely, given any set of univariate distribution functions $F_1,\ldots,F_N$ and any $N$-dimensional copula $C$, Eq.~\eqref{eq:sm-sklar} defines a valid joint CDF with marginals $F_i$.

For a continuous random variable $X_i$, the value
\begin{equation}
u_i=F_i(x_i)=\Pr(X_i\leq x_i),\qquad i=1,\ldots,N,
\label{eq:sm-percentile}
\end{equation}
intuitively represents the proportion of outcomes not exceeding $x_i$ in the marginal distribution of $X_i$. Thus, a copula can also be defined as the joint CDF of random variables $U_i$,
\begin{equation}
C(u_1,\ldots,u_N)
=\Pr(U_1\leq u_1,\ldots,U_N\leq u_N),
\label{eq:sm-copula-cdf}
\end{equation}
each of whose univariate marginals $U_i$ is uniformly distributed on $[0,1]$.

Equation~\eqref{eq:sm-sklar} expresses a beautiful idea: the statistical data of a whole composed of several parties can be elegantly decomposed into the statistical data of the individual parties (the marginals) and the statistical data characterizing the relational structure among them (the copula), and their combination uniquely recovers the joint distribution. Formally, we may write
\begin{equation}
\begin{aligned}
F &\longleftrightarrow
\bigl(\text{individual marginals},\\
&\hspace{2cm}\text{margin-free dependence structure}\bigr).
\end{aligned}
\label{eq:sm-parr}
\end{equation}

In principle, a discrete joint distribution can still be written in the form of Eq.~\eqref{eq:sm-sklar}. The difficulty is that the CDFs of discrete marginals are step functions and therefore do not cover the entire interval $[0,1]$.\footnote{For example, the range of the CDF of a binary variable may consist of only $\{0,p_i,1\}$.} Consequently, the joint distribution determines $C$ only on the discrete grid formed by the ranges of these marginals, not uniquely between the grid points. In other words, more than one continuous copula extension can be combined with the same discrete marginals to produce exactly the same discrete joint distribution. The key insight of Refs.~\cite{geenens2020,geenens2025} is that, in the discrete setting, rather than arbitrarily choosing a nonunique continuous copula, one can directly seek a discrete structure embodying the Sklar spirit, parallel to Eq.~\eqref{eq:sm-parr}:
\begin{equation}
\begin{gathered}
\text{joint probability table}\\
\longleftrightarrow
\bigl(\text{individual marginals},\\
\text{margin-free dependence structure}\bigr).
\end{gathered}
\label{eq:sm-parr2}
\end{equation}

In particular, Ref.~\cite{Lin:2026trf} gives a fully explicit realization of this abstract discrete Sklar structure in a special but exceptionally transparent sector: strictly positive $N$-binary distributions. More precisely, Ref.~\cite{Lin:2026trf} shows that this structure indeed exists, can be characterized completely, and, when recombined with arbitrary marginals, recovers exactly one joint table. We summarize these results in Table~\ref{tab:sm-sklar}.

\begin{table*}[t]
\caption{The questions addressed by the explicit Sklar-type structure for positive binary distributions in Ref.~\cite{Lin:2026trf}.}
\label{tab:sm-sklar}
\begin{ruledtabular}
\begin{tabular}{p{0.40\textwidth}p{0.43\textwidth}}
Question & Answer of Ref.~\cite{Lin:2026trf} \\
\hline
What does it mean to change only the individual parties? & Local positive reweighting \\
What remains invariant under this transformation? & All $J_S$ of second and higher orders \\
Are these invariants complete? & Yes; equal higher-order $J_S$ if and only if the tables belong to the same local-reweighting equivalence class \\
Can the relational data and marginals recover the whole? & Yes, and the reconstruction is unique \\
Is there a canonical representative? & Each equivalence class contains a unique table $\gamma_p$ with uniform marginals
\end{tabular}
\end{ruledtabular}
\end{table*}

We now describe $J_S$ more explicitly. For a joint distribution of binary variables $x_i\in\{-1,+1\}$, its log probability admits the unique expansion
\begin{equation}
\log p(x)=\sum_{S\subseteq[N]}J_S\chi_S(x),
\qquad \chi_S(x)=\prod_{i\in S}x_i.
\label{eq:sm-walsh}
\end{equation}
In this representation, local reweighting,
\begin{equation}
p(x)\longmapsto
\frac{p(x)\prod_i f_i(x_i)}
{\sum_y p(y)\prod_i f_i(y_i)},
\label{eq:sm-reweighting}
\end{equation}
changes only the constant term and the one-body coefficients $J_i$, whereas all coefficients $J_S$ involving at least two variables, $|S|\geq2$, remain invariant. Conversely, if two strictly positive binary probability tables have exactly the same higher-order coefficients, then they must differ only by such local reweighting.

Thus, Eq.~\eqref{eq:sm-parr} can be written more explicitly as
\begin{equation}
p\longleftrightarrow\bigl(\{p_i\},[p]\bigr),
\label{eq:sm-parr3}
\end{equation}
and
\begin{equation}
[p]=[q]\quad\Longleftrightarrow\quad
J_S(p)=J_S(q),\qquad |S|\geq2.
\label{eq:sm-corr}
\end{equation}
In Eq.~\eqref{eq:sm-parr3}, $\{p_i\}$ denotes the single-party marginals. The class $[p]$ consists of an entire family of probability tables related to one another by local reweighting. It expresses the idea that we allow the probability preferences of the individual parties to change while regarding the way they are combined as the same. It is therefore an abstract relational object, called the \emph{dependence nucleus} in Ref.~\cite{geenens2025}. Its continuous counterpart is the class of joint distributions sharing the same copula. In particular, let $\gamma_p$ denote the unique joint probability table in $[p]$ with uniform single-party marginals. The joint CDF $C_p$ obtained by cumulatively summing $\gamma_p$ can then be regarded as playing the role of a discrete copula. Equation~\eqref{eq:sm-corr}, meanwhile, shows that in the special strictly positive $N$-binary setting, an abstract dependence structure can be represented completely and explicitly by all higher-order $J_S$.

The significance of this representation is not merely that it provides a more convenient set of coordinates. The same $J_S$ already have familiar interpretations in different theories: in contingency-table theory, they are known as Walsh log-linear interactions and, in particular, can be equivalently extracted from conditional odds ratios. More interestingly, this framework also has a direct statistical-mechanical interpretation: the higher-order $J_S$ correspond precisely to the two-body and higher-order interaction couplings in finite-spin Gibbs systems. Thus, this explicit representation not only makes the general discrete Sklar structure fully concrete, but also directly unifies the corresponding objects in discrete copula theory, log-linear theory, and many-body couplings in classical statistical mechanics.

\subsection{Two Statements on Conditioning Higher-Order \texorpdfstring{$J_S$}{J_S}}

We now prove two statements closely related to our task of quantum-state reconstruction. Consider a strictly positive joint distribution $p(x)$ of $N$ binary variables $x_i\in\{-1,+1\}$. Fix the outcome $x_k=r$ of the $k$th variable, obtaining the conditional distribution $p_{\bar{k}\mid r}$ of the remaining $N-1$ variables.

\noindent\textbf{Proposition (i).---}
The dependence nucleus of the conditional distribution is determined solely by the original dependence nucleus $[p]$. More explicitly,
\begin{equation}
[p]=[q]\quad\Longrightarrow\quad
[p_{\bar{k}\mid r}]=[q_{\bar{k}\mid r}].
\label{eq:sm-S1}
\end{equation}

\noindent\textbf{Proposition (ii).---}
If $J_S$ are the log-linear coefficients of the original distribution and $J_T^{(k,r)}$ are the corresponding coefficients of the conditional distribution, then
\begin{equation}
\begin{gathered}
J_T^{(k,r)}=J_T+rJ_{T\cup\{k\}},\\
|T|\geq2,\qquad T\subseteq[N]\setminus\{k\}.
\end{gathered}
\label{eq:sm-S2}
\end{equation}

\noindent\textbf{Proof.---}
We first verify (i). In our discrete Sklar framework, two joint distributions $p,q$ belong to the same dependence nucleus, $[p]=[q]$, precisely when they differ only by local reweighting of the individual variables:
\begin{equation}
\begin{aligned}
q(x_1,\ldots,x_N)
&=\frac1Z p(x_1,\ldots,x_N)\\
&\quad\times\prod_{i=1}^{N}f_i(x_i),\qquad f_i>0.
\end{aligned}
\label{eq:sm-wei}
\end{equation}
Now condition on the last variable, for example by fixing $X_N=r$. The conditional distribution induced by $p$ is
\begin{equation}
p(x_1,\ldots,x_{N-1}\mid r)
=\frac{p(x_1,\ldots,x_{N-1},r)}{p(r)},
\label{eq:sm-pcond}
\end{equation}
while that induced by $q$ is
\begin{equation}
q(x_1,\ldots,x_{N-1}\mid r)
=\frac{q(x_1,\ldots,x_{N-1},r)}{q(r)}.
\label{eq:sm-qcond}
\end{equation}
Substituting Eq.~\eqref{eq:sm-wei}, we have
\begin{equation}
\begin{aligned}
&q(x_1,\ldots,x_{N-1},r)\\
&=\frac1Z p(x_1,\ldots,x_{N-1},r)
\left[\prod_{i=1}^{N-1}f_i(x_i)\right]f_N(r).
\end{aligned}
\label{eq:sm-wei-sub}
\end{equation}
Since $r$ is fixed, $f_N(r)$ is merely a common constant. Upon normalization, both it and $Z$ cancel, giving
\begin{equation}
\begin{aligned}
&q(x_1,\ldots,x_{N-1}\mid r)\\
&\quad\propto p(x_1,\ldots,x_{N-1}\mid r)
\prod_{i=1}^{N-1}f_i(x_i).
\end{aligned}
\label{eq:sm-rewei}
\end{equation}
Equation~\eqref{eq:sm-rewei} is exactly another local reweighting on the remaining $N-1$ variables. Therefore,
\begin{equation}
\begin{gathered}
[p]=[q]\quad\Longrightarrow\\
[p(\,\cdot\mid X_N=r)]
=[q(\,\cdot\mid X_N=r)].
\end{gathered}
\label{eq:sm-cond-nucleus}
\end{equation}

We next prove (ii). Recall the standard log-linear expansion,
\begin{equation}
\log p(x)=\sum_{S\subseteq[N]}J_S\prod_{i\in S}x_i.
\label{eq:sm-log}
\end{equation}
Substitute $x_N=r$. Clearly, the terms on the right-hand side of Eq.~\eqref{eq:sm-log} not containing $N$, namely $J_T\prod_{i\in T}x_i$, remain unchanged, whereas the terms containing $N$,
$J_{T\cup\{N\}}\bigl(\prod_{i\in T}x_i\bigr)x_N$,
become $rJ_{T\cup\{N\}}\prod_{i\in T}x_i$. Thus, in the conditional distribution, we obtain directly
\begin{equation}
J_T^{(N,r)}=J_T+rJ_{T\cup\{N\}}.
\label{eq:sm-cond-coeff}
\end{equation}
This formula tells us that conditioning does not discard higher-order dependence; rather, it folds the higher-order interactions involving the conditioned variable into lower-order conditional interactions among the remaining variables.

In particular, for $N=3$ with $C=r$ fixed, only two variables remain in the conditional table, whose relational data are completely characterized by the single coefficient $J_{AB}^{(C,r)}$. Using $4J_{AB}=\log\omega_{AB}$ for a bivariate binary table, Eq.~\eqref{eq:sm-S2} immediately gives
\begin{equation}
\log\omega_{AB\mid C=r}=4\bigl(J_{AB}+rJ_{ABC}\bigr).
\label{eq:sm-cond-odds}
\end{equation}

\section{Relational Data from Regular Born Tables Suffice}
\label{app:regular-born}

The main text shows that relational data from different measurement contexts of a quantum state can, remarkably, recover the local information originally discarded in the classical Sklar decomposition. There is, however, a subtlety: the coefficients $J_S$ used to characterize these relational data involve logarithms of probabilities, and hence are not automatically well defined for Born tables containing zero probabilities. Such cases can arise especially for pure states and other rank-deficient states. Fortunately, this does not obstruct our reconstruction. In fact, we need no information at all from these special Born tables: the relational data carried by the Born tables for which $J_S$ are well defined already suffice for the reconstruction developed in this work. We prove this below, paying particular attention to showing that these data also support our recursive reconstruction mechanism based on conditioning.

\subsection{Density of Regular Contexts}
\label{app:regular-density}

We call measurement contexts in which all probabilities are strictly positive \emph{regular contexts}. For one qubit, a Pauli measurement can be performed along any unit direction $\mathbf n$. Thus, all measurement directions for a qubit form the surface of the Bloch sphere:
\begin{equation}
S^2=\{\mathbf n\in\mathbb R^3:|\mathbf n|=1\}.
\label{eq:sm-sphere}
\end{equation}
More generally, for an arbitrary $N$-qubit density operator $\rho$, each party can choose its measurement direction independently, so the full space of measurement contexts is $\mathcal M=(S^2)^N$. For any outcome string $\mathbf s=(s_1,\ldots,s_N)$, its Born probability is
\begin{equation}
p_{\mathbf s}^{\rho}(\mathbf n)=
\operatorname{Tr}\!\left[
\rho\bigotimes_{i=1}^{N}
\frac{I+s_i\mathbf n_i\cdot\boldsymbol\sigma}{2}
\right].
\label{eq:sm-born-prob}
\end{equation}

What we first need to prove is that, for any fixed outcome string $\mathbf s$, the set of measurement directions satisfying $p_{\mathbf s}^{\rho}(\mathbf n)=0$ has measure zero. The proof requires only two key facts. The first is that a Born probability cannot vanish identically for all measurement directions. More precisely, averaging uniformly over all local measurement directions gives
\begin{equation}
\begin{aligned}
\int_{\mathcal M}p_{\mathbf s}^{\rho}(\mathbf n)\,d\mu(\mathbf n)
&=\operatorname{Tr}\!\left[
\rho\left(\frac I2\right)^{\!\otimes N}
\right]\\
&=2^{-N}>0.
\end{aligned}
\label{eq:sm-aver}
\end{equation}
Here $d\mu$ denotes the normalized uniform measure on $\mathcal M$. Notice that this result is completely independent of what state $\rho$ is! Equation~\eqref{eq:sm-aver} therefore tells us that, whether $\rho$ is pure, mixed, or highly special, every outcome string occurs with positive probability for some measurement directions. Hence no Born probability function can vanish identically. The second fact is that each Born probability is a real-analytic function of the measurement directions. A standard mathematical result states that, on a connected space, the zero set of a nonidentically vanishing real-analytic function has empty interior and measure zero. Since there are only finitely many cells in a probability table, and a finite union of measure-zero sets still has measure zero, the regular contexts for which all probabilities are strictly positive form an open, dense, full-measure set, denoted by
\begin{equation}
\mathcal R_{\rho}
=\left\{
\mathbf n\in\mathcal M:
p_{\mathbf s}^{\rho}(\mathbf n)>0
\ \text{for every }\mathbf s
\right\}.
\label{eq:sm-regular-set}
\end{equation}

Put simply, we find that for any quantum state, the relational data of almost all Born tables are well defined. More importantly, these contexts not only have full measure but are also dense throughout the measurement space. Full measure means that the discarded special contexts occupy a set of measure zero, ensuring that the relational data we use cover almost every measurement context. Density means that every discarded context can be approximated arbitrarily closely by regular contexts, allowing us to extend continuous, denominator-free identities obtained on regular contexts to all measurement directions, as we now show.\footnote{Openness means that a sufficiently small perturbation of any regular context remains regular. This follows from continuity and strict positivity of the probabilities.}

\subsection{The Bivariate Case}
\label{app:regular-bivariate}

We now use the density of regular contexts established in Sec.~\ref{app:regular-density} to demonstrate, first for two-qubit states, that even when certain special Born tables have no well-defined odds ratio, the remaining well-defined odds ratios already supply all the mathematical constraints required for the original reconstruction. Consider two candidate quantum states $\rho$ and $\sigma$. If their odds ratios coincide in all contexts that are regular for both, then
\begin{equation}
\frac{p_{++}^{\rho}p_{--}^{\rho}}
{p_{+-}^{\rho}p_{-+}^{\rho}}
=
\frac{p_{++}^{\sigma}p_{--}^{\sigma}}
{p_{+-}^{\sigma}p_{-+}^{\sigma}}.
\label{eq:sm-ratio}
\end{equation}
Notice that this equality currently holds only on regular contexts, where all four cell probabilities of each table are strictly positive. We now rewrite Eq.~\eqref{eq:sm-ratio} as
\begin{equation}
\begin{gathered}
p_{++}^{\rho}p_{--}^{\rho}
p_{+-}^{\sigma}p_{-+}^{\sigma}
=
p_{++}^{\sigma}p_{--}^{\sigma}
p_{+-}^{\rho}p_{-+}^{\rho}.
\end{gathered}
\label{eq:sm-cross}
\end{equation}
Clearly, the original equality \eqref{eq:sm-ratio} contains denominators and is therefore undefined at some special measurement directions. Equation~\eqref{eq:sm-cross}, however, has no denominators, so it remains meaningful even when a special direction produces a zero probability. More importantly, both sides of the equality are continuous functions of the measurement directions. Since we have just proved that regular contexts are dense, the equality established there must hold at every measurement direction.

What requires particular care here is that we extend the equality \emph{after clearing the denominators}; we do not extend an originally undefined odds ratio to Born tables containing zeros. If, for example, $p_{+-}=0$ at some special direction, the original odds ratio may remain undefined. There is no need to assign it a value. But the original two-qubit reconstruction proof ultimately uses the Bloch expression for Born probabilities, such as Eq.~\eqref{eq:pst}, to obtain algebraic constraints on the quantum-state parameters. We have now shown that the odds ratios on regular contexts alone yield exactly the same denominator-free algebraic identity, valid at all measurement directions. In other words, the algebraic argument that determines the quantum state retains all the constraints it needs. Thus, although we discard the relational data of some Born tables, we lose none of the algebraic constraints actually required for reconstruction.

\subsection{The Multipartite Case}
\label{app:regular-multipartite}

We now turn to the multipartite case. There are two steps, and they address different issues. First, we show that the relational data $J_S$ of general multipartite Born tables, just like the two-qubit odds ratio, also give rise to denominator-free identities that extend to all measurement contexts. Second, we show that even after conditioning, the remaining system still has sufficiently rich, well-defined relational data, and that the extension procedure from the first step remains applicable. We proceed in order.

The first step is entirely analogous to what we have just done in the bivariate case. In this procedure, the general multipartite $J_S$ are not fundamentally different from the two-qubit odds ratio. Recall that in the two-qubit case,
\begin{equation}
\omega=\frac{p_{++}p_{--}}
{p_{+-}p_{-+}}.
\label{eq:sm-odds-reminder}
\end{equation}
Although $\omega$ is undefined on certain zero-probability tables, equality of the odds ratios for two candidate states can be turned, by cross-multiplication, into an equation without denominators. Density of regular contexts and continuity of Born probabilities then extend that equation to all measurement directions. What about the general multipartite $J_S$? Recall their definition:
\begin{equation}
\begin{gathered}
J_S(p)=2^{-N}
\sum_{\mathbf s}
\chi_S(\mathbf s)\log p_{\mathbf s},\\
\chi_S(\mathbf s)=\prod_{i\in S}s_i.
\end{gathered}
\label{eq:sm-J-def}
\end{equation}
At first sight, these appear considerably more complicated than the two-qubit odds ratio. But notice that $\chi_S$ can take only the values $+1$ and $-1$. Thus, equality of the relational coordinates of two candidate states,
\begin{equation}
J_S(p^{\rho})=J_S(p^{\sigma}),
\label{eq:sm-J-equality}
\end{equation}
can in fact be exponentiated and rearranged directly into
\begin{equation}
\begin{gathered}
\prod_{\chi_S(\mathbf s)=+1}p_{\mathbf s}^{\rho}
\prod_{\chi_S(\mathbf s)=-1}p_{\mathbf s}^{\sigma}
=
\prod_{\chi_S(\mathbf s)=-1}p_{\mathbf s}^{\rho}
\prod_{\chi_S(\mathbf s)=+1}p_{\mathbf s}^{\sigma}.
\end{gathered}
\label{eq:sm-cross2}
\end{equation}
Thus, Eq.~\eqref{eq:sm-cross2} is again an identity involving only products of probabilities, with no logarithms or denominators! Once again, every Born probability depends continuously on the measurement directions. Therefore, whenever the original $J_S$ equality holds on the common regular contexts, their density ensures that Eq.~\eqref{eq:sm-cross2} automatically holds at all measurement directions. This proves that the general multipartite $J_S$ have exactly the same useful property as the two-qubit odds ratio: the algebraic identities derived from them can be extended. This completes the first step.

The second step asks, essentially: can things become worse after conditioning? This deserves even more attention than the first step, since it bears directly on our recursive reconstruction. Return, for example, to three qubits, and fix the measurement direction and outcome $r$ of $C$. We obtain the conditional Born table $p_{st\mid r}=p_{str}/p_r$. We now wish to apply the two-qubit theorem to reconstruct the two-qubit density operator that uniformly represents this family of conditional Born distributions. Naturally, there are two concerns. The first is whether conditional Born tables might contain so many zero probabilities that there are not enough well-defined odds ratios. The answer is no. As long as the fixed event $C=r$ has positive probability, the family of conditional Born probabilities of the other two parties can be represented by a valid two-qubit density operator $\widehat{\tau}_{AB\mid C,r}$. The open, dense, full-measure result proved at the beginning of this section holds for \emph{any} valid density operator. It therefore applies to $\widehat{\tau}_{AB\mid C,r}$ as well. Importantly, we are not simply inferring the density of the remaining two-qubit regular contexts, after fixing $C$, from the density of the original three-qubit regular contexts. Rather, we apply the original regular-context theorem afresh to the unified density-operator representation of the conditional probability family, thereby obtaining the conditional system's own open dense set.

Our second concern is that, even if the conditional system has sufficiently rich regular contexts, we must still verify that equality of the relational data of two candidate parent states really guarantees equality of the relational data of their conditional systems. But this is already guaranteed by the conditioning inheritance formula \eqref{eq:sm-S2}. That formula tells us that the relational data of a conditional table come entirely from the relational data of the original Born table. Therefore, when the original Born tables are strictly positive, if two candidate parent states have identical higher-order $J_S$, their conditional relational coordinates $J_T^{(k,r)}$ must also coincide. There is, however, one more subtlety: what if a particular fixed conditioning direction happens to produce a zero probability in the original Born table, so that Eq.~\eqref{eq:sm-S2} cannot be used directly? We need only apply the denominator-free extension procedure once more. More explicitly,
\begin{equation}
\begin{aligned}
J_T^{(k,r)}
&=2^{-(N-1)}
\sum_{\mathbf s_{\bar k}}
\chi_T(\mathbf s_{\bar k})\times
\log p_{\mathbf s_{\bar k},r},
\end{aligned}
\label{eq:sm-cond-J-log}
\end{equation}
where the conditioning normalization factor $p_r$ has already canceled automatically. Thus, equality of the conditional $J_T^{(k,r)}$ for two candidate states can likewise be rearranged into
\begin{equation}
\begin{gathered}
\prod_{\chi_T=+1}
p_{\mathbf s_{\bar k},r}^{\rho}
\prod_{\chi_T=-1}
p_{\mathbf s_{\bar k},r}^{\sigma}
=
\prod_{\chi_T=-1}
p_{\mathbf s_{\bar k},r}^{\rho}
\prod_{\chi_T=+1}
p_{\mathbf s_{\bar k},r}^{\sigma}.
\end{gathered}
\label{eq:sm-cross3}
\end{equation}
Here each product runs over the remaining outcome strings $\mathbf s_{\bar k}$ for which $\chi_T(\mathbf s_{\bar k})$ has the indicated sign. This is again a denominator-free identity in the original Born probabilities. We can first establish Eq.~\eqref{eq:sm-cross3} on the common regular contexts of the original systems and then extend it to all measurement directions using density and continuity. Thus, even after fixing a special conditioning direction, provided the corresponding conditioning event has positive probability for both candidate states, we can still use Eq.~\eqref{eq:sm-cross3}. Moreover, on the regular contexts of the remaining system, we can take logarithms of Eq.~\eqref{eq:sm-cross3} in reverse and conclude that the conditional relational data of the two candidates agree. We have therefore proved that the all-context dependence data needed by the lower-body reconstruction theorem are indeed supplied by the regular relational data of the original parent states.

In short, we have proved that zero probabilities do not disrupt the recursive procedure.

\section{Supplementary Technical Details}
\label{app:technical-details}

\subsection{Three-Qubit State Reconstruction}
\label{app:three-qubit-details}

\subsubsection{Rigidity of Conditional Representations}
\label{app:conditional-rigidity}

In this section, we address the technical details promised in the three-qubit reconstruction section.

We first prove the following: if the conditional representation on $AB$ is a product for every direction along which $C$ is measured, then either $A$ or $B$ must be isolated in the three-qubit parent state, namely,
\begin{equation}
\rho_{ABC}=\rho_A\otimes\rho_{BC}
\quad\text{or}\quad
\rho_{ABC}=\rho_{AC}\otimes\rho_B.
\end{equation}
We may call this the rigidity of conditional representations.

Mathematically, however, it turns out that we need only prove the following equivalent lemma.

\medskip
\noindent\textbf{Lemma S3.1 (Rigidity of conditional representations).---}
Let $\rho_{ABC}$ be an arbitrary three-qubit density operator, and fix the measurement outcome $r=\pm1$ of qubit $C$. Suppose there exists a nonempty open region $\Omega$ on the Bloch sphere such that, for every $\mathbf n_C\in\Omega$, the corresponding conditioning probability satisfies $p_r^\rho(\mathbf n_C)>0$ and the two-qubit representation $\widehat\tau_{AB\mid C,r}^\rho(\mathbf n_C)$ is a product. Then the parent state necessarily satisfies
\begin{equation}
\rho_{ABC}=\rho_A\otimes\rho_{BC}
\quad\text{or}\quad
\rho_{ABC}=\rho_{AC}\otimes\rho_B.
\label{eq:sm-s31-lemma}
\end{equation}
\medskip

\noindent\textit{Proof.---}
We start from \MainEq{eq:con} of the main text, reproduced here:
\begin{equation}
\begin{aligned}
p_r^\rho(\mathbf n_C)\widehat\tau_{AB\mid C,r}^\rho(\mathbf n_C)
=\frac12\biggl(M_0^{AB}
+r\sum_{\mu=x,y,z}n_{C,\mu}M_\mu^{AB}\biggr).
\end{aligned}
\label{eq:sm-s31-parent}
\end{equation}
Expand each of the four fixed operators $M_\nu^{AB}$ in the local Pauli basis:
\begin{equation}
M_\nu^{AB}=\sum_{\alpha,\beta=0,x,y,z}
(C_\nu)_{\alpha\beta}\,
\sigma_\alpha^A\otimes\sigma_\beta^B,
\quad \nu=0,x,y,z.
\label{eq:sm-s31-pauli}
\end{equation}
Here $(C_\nu)_{\alpha\beta}\in\mathbb R$ are the expansion coefficients. In other words, $C_0,C_x,C_y,C_z$ are four fixed real $4\times4$ coefficient matrices. Substituting Eq.~\eqref{eq:sm-s31-pauli} into Eq.~\eqref{eq:sm-s31-parent}, we obtain
\begin{equation}
\begin{aligned}
&p_r^\rho(\mathbf n_C)\widehat\tau_{AB\mid C,r}^\rho(\mathbf n_C)\\
&\quad=\sum_{\alpha,\beta}\left[\frac12\left(C_0+
 r\sum_\mu n_{C,\mu}C_\mu\right)\right]_{\alpha\beta}
\sigma_\alpha^A\otimes\sigma_\beta^B.
\end{aligned}
\label{eq:sm-s31-coeff-form}
\end{equation}
Define $C(\mathbf n_C)$ as the coefficient matrix of the entire operator on the right-hand side of Eq.~\eqref{eq:sm-s31-parent}, obtained by combining the four $C_\nu$ according to the measurement direction:
\begin{equation}
C(\mathbf n_C):=\frac12\left(C_0+
 r\sum_{\mu=x,y,z}n_{C,\mu}C_\mu\right).
\label{eq:sm-s31-affine}
\end{equation}
Thus, the Pauli coefficient matrices on the right-hand side of Eq.~\eqref{eq:sm-s31-parent} form an affine family in the measurement direction. Since the Pauli coefficient matrix of a product operator has rank one, and scalar multiplication does not change this property, the lemma's hypothesis implies
\begin{equation}
\operatorname{rank}C(\mathbf n_C)\leq1,
\qquad \mathbf n_C\in\Omega.
\end{equation}
Consequently, all $2\times2$ minors of $C(\mathbf n_C)$ vanish on $\Omega$. Each minor is a polynomial of degree at most two in $\mathbf n_C$. By analyticity on the Bloch sphere, these polynomials vanish identically on the entire unit sphere and hence must have the form
\begin{equation}
q(\mathbf n_C)=\lambda(1-|\mathbf n_C|^2).
\label{eq:sm-s31-sphere-poly}
\end{equation}

We now use the determinant structure of the minors to exclude the possibility $\lambda\neq0$. Homogenizing $q$ gives a quadratic form in four real variables,
\begin{equation}
\begin{aligned}
Q(t,\mathbf n_C)
&=\det\left(tD_0+r\sum_\mu n_{C,\mu}D_\mu\right)\\
&=\lambda(t^2-|\mathbf n_C|^2),
\end{aligned}
\label{eq:sm-s31-quadratic}
\end{equation}
where $D_\nu$ are the corresponding $2\times2$ submatrices. Since $Q$ is a $2\times2$ determinant, its positive and negative inertia indices are each at most two. However, when $\lambda\neq0$, the right-hand side of Eq.~\eqref{eq:sm-s31-quadratic} has signature $(1,3)$ or $(3,1)$, a contradiction. Therefore $\lambda=0$.

This shows that all minors in fact vanish as polynomial identities, so that
\begin{equation}
\operatorname{rank}\left(t_0C_0+
\sum_{\mu=x,y,z}t_\mu C_\mu\right)\leq1,
\quad \forall\,t_\nu\in\mathbb R.
\label{eq:sm-s31-span-rank}
\end{equation}
In other words, every linear combination of the four operators $M_\nu^{AB}$ is a product operator or the zero operator.

We now use a structural property of rank-one linear subspaces: a linear space consisting only of rank-one matrices and the zero matrix must have a common left factor or a common right factor. Otherwise, if two matrices had linearly independent factors on both sides, their sum would have rank two, contradicting the hypothesis.

There are therefore two possibilities:
\begin{equation}
M_\nu^{AB}=X_A\otimes Y_{\nu,B},
\qquad \nu=0,x,y,z,
\label{eq:sm-s31-common-left}
\end{equation}
or
\begin{equation}
M_\nu^{AB}=X_{\nu,A}\otimes Y_B,
\qquad \nu=0,x,y,z.
\label{eq:sm-s31-common-right}
\end{equation}
Finally, substituting Eq.~\eqref{eq:sm-s31-common-left} into \MainEq{eq:exp} of the main text gives
\begin{equation}
\begin{aligned}
\rho_{ABC}
&=\frac12\left(M_0^{AB}\otimes I_C+
\sum_\mu M_\mu^{AB}\otimes\sigma_\mu^C\right)\\
&=X_A\otimes\frac12\left(Y_{0,B}\otimes I_C+
\sum_\mu Y_{\mu,B}\otimes\sigma_\mu^C\right)\\
&\equiv X_A\otimes R_{BC}.
\end{aligned}
\label{eq:sm-s31-parent-factor}
\end{equation}
Taking partial traces yields
\begin{equation}
\rho_A=(\operatorname{Tr}R_{BC})X_A,
\qquad \rho_{BC}=(\operatorname{Tr}X_A)R_{BC}.
\end{equation}
Hence, using $\operatorname{Tr}\rho_{ABC}=1$,
\begin{equation}
\rho_A\otimes\rho_{BC}
=X_A\otimes R_{BC}=\rho_{ABC}.
\end{equation}
Thus $A$ is isolated. Equation~\eqref{eq:sm-s31-common-right} similarly gives $\rho_{ABC}=\rho_{AC}\otimes\rho_B$. This proves the lemma.

\subsubsection{Local Consistency of the Reconstruction Branches}
\label{app:branch-consistency}

Suppose two three-qubit parent states $\rho_{ABC}$ and $\sigma_{ABC}$ have identical all-context dependence data. Fix the measurement outcome $r$ of $C$ while varying its measurement direction $\mathbf n_C$. We already know that, for directions where the two-qubit reconstruction theorem applies, their conditional representations must satisfy one of two relations:

First, the two conditional representations coincide:
\begin{equation}
\widehat\tau^\sigma(\mathbf n_C)=
\widehat\tau^\rho(\mathbf n_C).
\tag{I}\label{eq:sm-s31-identity-branch}
\end{equation}
Second, the two conditional representations differ by a simultaneous double spin flip:
\begin{equation}
\widehat\tau^\sigma(\mathbf n_C)=
\mathcal F_A\mathcal F_B\bigl(\widehat\tau^\rho(\mathbf n_C)\bigr).
\tag{II}\label{eq:sm-s31-flip-branch}
\end{equation}
Notice, however, that the two-qubit theorem tells us only, for each measurement direction separately, that ``(I) or (II) holds.'' It does not tell us whether (I) holds for every direction, whether (II) holds for every direction, or whether some directions satisfy (I) and others satisfy (II). It is the last possibility that requires care. Nevertheless, we can prove that there exists at least one nonempty open region on which one branch holds consistently. We now give the details.

Consider a nonempty open region $U$ in which every direction satisfies either (I) or (II). Could (I) hold throughout $U$? Certainly! In that case we are already done: take $U$ itself as the open region on which a single branch holds. Now consider the other possibility, that (I) does not hold throughout $U$. Then there must be at least one direction $\mathbf n_*$ for which
\begin{equation}
\widehat\tau^\sigma(\mathbf n_*)\neq
\widehat\tau^\rho(\mathbf n_*).
\end{equation}
At this direction, (I) is ruled out. Since the two-qubit theorem guarantees that at least one of (I) and (II) holds, (II) must hold at $\mathbf n_*$. Now note that both conditional representations depend continuously on the measurement direction. Thus, if they are unequal at $\mathbf n_*$, they remain unequal at all sufficiently nearby directions. In other words, continuity guarantees a small nonempty open neighborhood $V$ of $\mathbf n_*$ on which
\begin{equation}
\widehat\tau^\sigma(\mathbf n_C)\neq
\widehat\tau^\rho(\mathbf n_C),
\qquad \forall\,\mathbf n_C\in V.
\end{equation}
Throughout $V$, therefore, the identity branch (I) is impossible. But the two-qubit reconstruction theorem again guarantees that either (I) or (II) must hold at every direction. Consequently, only the second branch remains:
\begin{equation}
\widehat\tau^\sigma(\mathbf n_C)=
\mathcal F_A\mathcal F_B\bigl(\widehat\tau^\rho(\mathbf n_C)\bigr),
\quad\forall\,\mathbf n_C\in V.
\end{equation}
In short, in either case we can always find a nonempty open region of conditioning directions on which the identity branch or the simultaneous double-spin-flip branch holds consistently.

\subsubsection{Coefficient Comparison  and Degenerate Cases}
\label{app:coefficient-rigidity}

We already know that, on some nonempty open region of measurement directions, the conditional representations of the two candidate parent states coincide:
\begin{equation}
\widehat\tau_{AB\mid C,r}^\sigma(\mathbf n_C)=
\widehat\tau_{AB\mid C,r}^\rho(\mathbf n_C).
\end{equation}
On the other hand, we have \MainEq{eq:squa}, reproduced here:
\begin{equation}
\begin{aligned}
&p_r^\sigma(\mathbf n_C)\left(M_0^{AB}+
 r\sum_\mu n_{C,\mu}M_\mu^{AB}\right)\\
&\quad=p_r^\rho(\mathbf n_C)\left(N_0^{AB}+
 r\sum_\mu n_{C,\mu}N_\mu^{AB}\right).
\end{aligned}
\label{eq:sm-s31-squa1}
\end{equation}
Our task is to prove rigorously that, apart from the degenerate case in which $C$ is isolated, one must have
\begin{equation}
p_r^\sigma(\mathbf n_C)=p_r^\rho(\mathbf n_C),
\end{equation}
and hence that the four $M_\nu^{AB}$ and $N_\nu^{AB}$ coincide term by term.

First, Eq.~\eqref{eq:sm-s31-squa1} tells us that, on the identity branch, the two operator families generated by the parent-state Pauli coefficients are proportional throughout a nonempty open region of the Bloch sphere. To use this constraint rigorously, define real-linear maps
\begin{equation}
\begin{aligned}
L_\rho(t,\mathbf v)&=tM_0^{AB}+
 r\sum_\mu v_\mu M_\mu^{AB},\\
L_\sigma(t,\mathbf v)&=tN_0^{AB}+
 r\sum_\mu v_\mu N_\mu^{AB}.
\end{aligned}
\label{eq:sm-s31-Lmaps}
\end{equation}
Then Eq.~\eqref{eq:sm-s31-squa1} implies
\begin{equation}
L_\rho(1,\mathbf n_C)\wedge
L_\sigma(1,\mathbf n_C)=0,
\quad\mathbf n_C\in U,
\end{equation}
where the exterior product expresses linear dependence in the real vector space of Hermitian operators. In any real operator basis, each component of this identity is a $2\times2$ determinant of linear polynomials. We may therefore invoke directly the Bloch-sphere quadratic-determinant lemma already proved in Sec.~\ref{app:conditional-rigidity}: if such a determinant vanishes throughout a nonempty open region of the sphere, it vanishes as a polynomial identity. Homogenizing gives
\begin{equation}
L_\rho(z)\wedge L_\sigma(z)=0,
\qquad \forall\,z\in\mathbb R^4.
\label{eq:sm-s31-wedge}
\end{equation}
We now use the following elementary linear-algebra lemma: if two real-linear maps have linearly dependent outputs for every common input, and one of the maps has rank at least two, then the two maps differ only by a common constant factor. Thus, provided that
\begin{equation}
\operatorname{rank}L_\rho\geq2,
\end{equation}
there exists a constant $c$ such that
\begin{equation}
L_\sigma=cL_\rho.
\end{equation}
This rank condition is ensured precisely by the assumption that no qubit is isolated. Indeed, if $\operatorname{rank}L_\rho=1$, then, since $M_0^{AB}=\rho_{AB}\neq0$, necessarily
\begin{equation}
M_\mu^{AB}=a_\mu M_0^{AB}.
\end{equation}
Substituting this into \MainEq{eq:exp} of the main text yields
\begin{equation}
\begin{aligned}
\rho_{ABC}
&=\rho_{AB}\otimes\frac12\left(I_C+
\sum_\mu a_\mu\sigma_\mu^C\right)\\
&=\rho_{AB}\otimes\rho_C,
\end{aligned}
\end{equation}
contrary to the assumption.

Therefore,
\begin{equation}
N_\nu^{AB}=cM_\nu^{AB},
\qquad \nu=0,x,y,z.
\end{equation}
Normalization of the parent states gives $c=1$. This rigorously establishes \MainEq{eq:nm} of the main text, and \MainEq{eq:con} then gives \MainEq{eq:wise}.

For the simultaneous double-spin-flip branch, one need only replace $L_\rho$ above by $\mathcal F_A\mathcal F_B\circ L_\rho$. This map is fixed, invertible, trace-preserving, and real-linear, so the same argument applies.

\subsection{N-Qubit State Reconstruction}
\label{app:n-qubit-details}

\subsubsection{Irreducibility under Generic Conditioning}
\label{app:generic-conditioning}

Here we supplement the proof of a statement used in the $N$-qubit reconstruction in the main text: if an $N$-qubit parent state $\rho$ admits no nontrivial tensor-product cut, then, after fixing the measurement outcome $r$ of the $k$th qubit, the unified density-operator representation of the conditional Born distributions of the remaining $N-1$ qubits remains irreducible for generic measurement directions $\mathbf n_k$.

Let $\bar k$ denote all subsystems other than qubit $k$. From \MainEq{eq:N-con} of the main text, reproduced here,
\begin{equation}
\begin{aligned}
&p_r^\rho(\mathbf n_k)
\widehat\tau_{\bar k\mid k,r}^\rho(\mathbf n_k)\\
&\quad=\frac12\left[M_0^{\bar k}+
 r\sum_{\mu=x,y,z}n_{k,\mu}M_\mu^{\bar k}\right].
\end{aligned}
\label{eq:sm-s32-parent}
\end{equation}
Since the parent state is irreducible, its single-qubit reduced state $\rho_k$ must have full rank; otherwise qubit $k$ itself would be a tensor-product factor. Thus $p_r^\rho(\mathbf n_k)>0$ for all unit directions, and product factorization of the conditional representation is equivalent to product factorization of the operator on the right-hand side of Eq.~\eqref{eq:sm-s32-parent}. Now fix any nontrivial partition of the remaining system,
\begin{equation}
\bar k=S\sqcup T,
\qquad S,T\neq\varnothing.
\end{equation}
Choose real bases for the Hermitian operator spaces of $S$ and $T$, respectively, and expand the operator in square brackets on the right-hand side of Eq.~\eqref{eq:sm-s32-parent} into a coefficient matrix
\begin{equation}
C(\mathbf n_k)=C_0+
 r\sum_{\mu=x,y,z}n_{k,\mu}C_\mu.
\label{eq:sm-s32-affine}
\end{equation}
Here the $C_\nu$ correspond to the four Pauli expansion coefficients $M_\nu^{\bar k}$ of the parent state. In this matrix representation, the conditional operator factors across $S\mid T$ if and only if $C(\mathbf n_k)$ has rank one. Suppose there exists a nonempty open region $U$ of the Bloch sphere such that the conditional representation factors across $S\mid T$ for every $\mathbf n_k\in U$. Then all $2\times2$ minors of $C(\mathbf n_k)$ vanish on $U$. Each such minor is a polynomial of degree at most two in $\mathbf n_k$, with the determinant structure of two linear expressions. We may therefore invoke directly the Bloch-sphere quadratic-determinant lemma proved in Sec.~\ref{app:conditional-rigidity}: any such minor vanishing on a nonempty open region of the unit sphere must vanish identically as a polynomial. Homogenizing Eq.~\eqref{eq:sm-s32-affine}, we find that, for arbitrary real $t,v_x,v_y,v_z$, the matrix
\begin{equation}
tC_0+r\sum_{\mu=x,y,z}v_\mu C_\mu
\label{eq:sm-s32-linear-span}
\end{equation}
has rank at most one. Equivalently, the four matrices $C_0,C_x,C_y,C_z$ span a real-linear subspace consisting entirely of matrices of rank at most one. Applying again the rank-one linear-subspace structure result from Sec.~\ref{app:conditional-rigidity}, such a matrix subspace must have a common left factor or a common right factor. It follows immediately that the four operators $M_\nu^{\bar k}$ have one of the two forms
\begin{equation}
M_\nu^{\bar k}=X_S\otimes Y_\nu^T,
\qquad \nu=0,x,y,z,
\label{eq:sm-s32-left}
\end{equation}
or
\begin{equation}
M_\nu^{\bar k}=X_\nu^S\otimes Y_T,
\qquad \nu=0,x,y,z.
\label{eq:sm-s32-right}
\end{equation}
Here $X_S$ in the first case is independent of $\nu$, while $Y_T$ in the second case is independent of $\nu$. Substituting this mathematical conclusion back into the parent-state Pauli expansion, Eq.~\eqref{eq:sm-s32-left}, for example, gives
\begin{equation}
\begin{aligned}
\rho
&=\frac12\left[M_0^{\bar k}\otimes I_k+
\sum_\mu M_\mu^{\bar k}\otimes\sigma_\mu^k\right]\\
&=X_S\otimes\frac12\left[Y_0^T\otimes I_k+
\sum_\mu Y_\mu^T\otimes\sigma_\mu^k\right].
\end{aligned}
\label{eq:sm-s32-factorized-parent}
\end{equation}
By normalization of $\rho$ and taking partial traces over the two sides, the normalized forms of the two factors in Eq.~\eqref{eq:sm-s32-factorized-parent} must be $\rho_S$ and $\rho_{T\cup\{k\}}$, respectively. Therefore,
\begin{equation}
\rho=\rho_S\otimes\rho_{T\cup\{k\}}.
\label{eq:sm-s32-product-left}
\end{equation}
Equation~\eqref{eq:sm-s32-right} similarly gives
\begin{equation}
\rho=\rho_{S\cup\{k\}}\otimes\rho_T.
\label{eq:sm-s32-product-right}
\end{equation}
Both possibilities imply a nontrivial tensor-product cut of the parent state, contradicting its irreducibility. Thus, for any fixed partition $S\mid T$, the set of measurement directions for which the conditional representation factors across that partition cannot contain a nonempty open region of the sphere.

Finally, note that the remaining $N-1$ qubits have only finitely many nontrivial bipartitions. For each fixed partition, the corresponding set of bad directions is the common zero set of the $2\times2$ minors in Eq.~\eqref{eq:sm-s32-affine} and is therefore closed; by the argument above, it has empty interior and is hence nowhere dense. A finite union of closed nowhere-dense sets is still nowhere dense. Hence the set of all measurement directions for which the conditional representation admits some nontrivial tensor-product cut is a closed set with empty interior, whose complement is a nonempty open dense set. Moreover, for every fixed partition at least one minor is not identically zero on the sphere. By the standard zero-set property of nonzero real-analytic functions, the corresponding bad set has measure zero on the sphere. This remains true after taking the finite union over partitions. Therefore, for generic measurement directions $\mathbf n_k$,
\begin{equation}
\widehat\tau_{\bar k\mid k,r}^\rho(\mathbf n_k)
\ \text{admits no nontrivial tensor-product cut}.
\label{eq:sm-s32-generic-irred}
\end{equation}
This rigorously establishes the irreducibility under generic conditioning required for $N$-qubit reconstruction in the main text and ensures that the induction hypothesis can be applied on a nonempty open set of regular conditioning directions.

\subsubsection{Dimension Independence of Conditioning-Probability Recovery}
\label{app:dimension-independent}

We now supplement the proof that the conditioning-probability recovery used in the $N$-qubit reconstruction in the main text does not depend on the Hilbert-space dimension of the retained system. Let $N\geq4$, let $\rho$ be an irreducible $N$-qubit parent state, and let $\sigma$ be a candidate state with the same all-context dependence data. Fix the measurement outcome $r$ of qubit $k$. By the preceding proof of irreducibility under generic conditioning and the induction hypothesis for $(N-1)$-qubit reconstruction, on a nonempty open region $U$ of the Bloch sphere we have
\begin{equation}
\widehat\tau_{\bar k\mid k,r}^\sigma(\mathbf n_k)
=\widehat\tau_{\bar k\mid k,r}^\rho(\mathbf n_k),
\qquad \mathbf n_k\in U.
\label{eq:sm-s3221}
\end{equation}
Here $\bar k$ denotes the remaining $N-1$ qubits. Following \MainEq{eq:N-con} of the main text, denote the Pauli expansion coefficients of the two parent states with respect to qubit $k$ by $M_\nu^{\bar k}$ and $N_\nu^{\bar k}$, respectively. Eliminating the common normalized conditional representation in Eq.~\eqref{eq:sm-s3221}, we obtain
\begin{equation}
\begin{aligned}
&p_r^\sigma(\mathbf n_k)\left[M_0^{\bar k}+
 r\sum_{\mu=x,y,z}n_{k,\mu}M_\mu^{\bar k}\right]\\
&\quad=p_r^\rho(\mathbf n_k)\left[N_0^{\bar k}+
 r\sum_{\mu=x,y,z}n_{k,\mu}N_\mu^{\bar k}\right].
\end{aligned}
\label{eq:sm-s3222}
\end{equation}
This has exactly the same mathematical structure as \MainEq{eq:squa} in the three-qubit reconstruction section of the main text. Indeed, although the four Pauli expansion coefficients now act on a larger Hilbert space, their dependence on the measurement direction still contains only a constant term and terms of first order. Accordingly, define two real-linear maps
\begin{equation}
\begin{aligned}
L_\rho(t,\mathbf v)&=tM_0^{\bar k}+
 r\sum_\mu v_\mu M_\mu^{\bar k},\\
L_\sigma(t,\mathbf v)&=tN_0^{\bar k}+
 r\sum_\mu v_\mu N_\mu^{\bar k},
\end{aligned}
\label{eq:sm-s32-Lmaps}
\end{equation}
where
\begin{equation}
L_\rho,L_\sigma:\mathbb R^4\longrightarrow
\operatorname{Herm}(\mathcal H_{\bar k}).
\end{equation}
Equation~\eqref{eq:sm-s3222} means that, for every $\mathbf n_k\in U$, the outputs of the two maps at $(1,\mathbf n_k)$ are linearly dependent. By the Bloch-sphere quadratic-determinant lemma proved in Sec.~\ref{app:conditional-rigidity}, this constraint can be homogenized into
\begin{equation}
L_\rho(z)\wedge L_\sigma(z)=0,
\qquad \forall\,z\in\mathbb R^4.
\label{eq:sm-s3223}
\end{equation}
The key point is that both the sphere-determinant lemma and the proportionality-rigidity result for pointwise linearly dependent maps from Sec.~\ref{app:coefficient-rigidity} are independent of the dimension of the codomain. Therefore, provided that
\begin{equation}
\operatorname{rank}L_\rho\geq2,
\end{equation}
we may apply the latter linear-algebra lemma directly and obtain
\begin{equation}
L_\sigma=cL_\rho,
\label{eq:sm-s3224}
\end{equation}
where $c$ is a real constant independent of $z$.

This rank condition is guaranteed precisely by the irreducibility of the parent state. Indeed, if $\operatorname{rank}L_\rho=1$, then since $M_0^{\bar k}=\rho_{\bar k}\neq0$, one must have
\begin{equation}
M_\mu^{\bar k}=a_\mu M_0^{\bar k},
\qquad \mu=x,y,z.
\end{equation}
Substituting into the parent-state Pauli expansion yields
\begin{equation}
\begin{aligned}
\rho
&=\frac12\left[M_0^{\bar k}\otimes I_k+
\sum_\mu M_\mu^{\bar k}\otimes\sigma_\mu^k\right]\\
&=\rho_{\bar k}\otimes\frac12\left(I_k+
\sum_\mu a_\mu\sigma_\mu^k\right)\\
&=\rho_{\bar k}\otimes\rho_k.
\end{aligned}
\end{equation}
This means that qubit $k$ was isolated from the outset, contradicting the irreducibility of the parent state. Hence $\operatorname{rank}L_\rho\geq2$, and Eq.~\eqref{eq:sm-s3224} necessarily holds.

Comparing the four Pauli expansion coefficients then immediately gives
\begin{equation}
N_\nu^{\bar k}=cM_\nu^{\bar k},
\qquad \nu=0,x,y,z.
\label{eq:sm-s3225}
\end{equation}
Since these four coefficients completely determine the parent state, $\sigma=c\rho$. Normalization of the two parent states then requires $c=1$. Thus,
\begin{equation}
p_r^\sigma(\mathbf n_k)=p_r^\rho(\mathbf n_k),
\end{equation}
and
\begin{equation}
N_\nu^{\bar k}=M_\nu^{\bar k},
\qquad\nu=0,x,y,z.
\end{equation}
By uniqueness of the Pauli expansion, we finally obtain
\begin{equation}
\sigma=\rho.
\end{equation}
This proves that the conditioning-probability recovery mechanism required for $N$-qubit reconstruction in the main text is independent of the dimension of the remaining system, completing the induction step.

\bibliography{references}
\end{document}